\documentclass[10pt, conference, a4paper]{IEEEtran}
\usepackage[letterpaper, left = 0.68in, right =0.68in, top = 0.7in, bottom = 1.042in]{geometry}

\def\review{0}%
\def\arxivdisclaimer{1}%

\usepackage[noadjust]{cite}
\usepackage{amsmath}
\usepackage{amssymb}
\usepackage{amsthm}
\usepackage{amsfonts}
\usepackage{graphicx}
\usepackage{textcomp}
\usepackage{mathtools}
\usepackage{adjustbox}
\usepackage{balance}
\usepackage{verbatim}
\usepackage{bm}
\usepackage{tikz}
\usepackage{titlesec}
\usetikzlibrary{arrows, fit, matrix, positioning, shapes, backgrounds, spy, shapes.geometric, arrows}
\usepackage[dvipsnames]{xcolor}
\usepackage{textcase}
\usepackage[tablename=TABLE,font=small]{caption}

\usepackage{paralist}
\usepackage{supertabular}    
\usepackage{array} 
\usepackage[linesnumbered,ruled]{algorithm2e}
\usepackage{algorithmic}
\usepackage[acronym]{glossaries}

\if\review1
\usepackage{todonotes}
\else
\usepackage[disable]{todonotes}
\fi

\usepackage{nohyperref}

\usepackage{numprint}     
\usepackage{makecell}   
\usepackage{colortbl}
\usepackage{enumitem}            
\usepackage{longtable} 
\usepackage{wrapfig}
\usepackage{booktabs}
\usepackage{graphics}
\usepackage{pgfplots}
\usepgfplotslibrary{groupplots}
\usepackage{pgfplotstable}
\usepackage{subcaption}
\pgfplotsset{compat=1.18} 

\usepackage{tabu}

\usepackage[utf8]{inputenc}

\usepackage{float}

\usepackage{psfrag}

\newcommand{\fakepar}[1]{\vspace{0mm}\noindent \textbullet \hspace{1mm}\textbf{\textit{#1.}}}

\newcommand{\egc}{e.g., }
\newcommand{\iec}{i.e., }

\newtheorem{remark}{Remark}
\newtheorem*{remark*}{Remark}

\def\BibTeX{{\rm B\kern-.05em{\sc i\kern-.025em b}\kern-.08em
    T\kern-.1667em\lower.7ex\hbox{E}\kern-.125emX}}

\DeclareCaptionTextFormat{up}{\MakeTextUppercase{#1}}

\let\oldtabular\tabular
\renewcommand{\tabular}{\small\oldtabular}

\newcolumntype{?}{!{\vrule width 1pt}}

\newcommand\blfootnote[1]{%
  \begingroup
  \renewcommand\thefootnote{}\footnote{#1}%
  \addtocounter{footnote}{-1}%
  \endgroup
}

\definecolor{mittelblau}{RGB}{0, 126, 198}
\definecolor{violettblau}{cmyk}{0.9, 0.6, 0, 0}
\definecolor{rot}{RGB}{238, 28 35}
\definecolor{apfelgruen}{RGB}{140, 198, 62}
\definecolor{gelb}{RGB}{1, 221, 0}
\definecolor{orange}{RGB}{244, 111, 33}
\definecolor{pink}{RGB}{237, 0, 140}
\definecolor{lila}{RGB}{128, 10, 145}
\definecolor{hellgrau}{RGB}{224, 224, 224}
\definecolor{mittelgrau}{RGB}{128, 128, 128}
\definecolor{dunkelgrau}{RGB}{80,80,80}
\definecolor{anthrazit}{RGB}{19, 31, 31}
\definecolor{darkgreen}{RGB}{0.125,0.5,0.169}
\definecolor{ahmedyellow}{RGB}{204,153,0}

\newacronym{3gpp}{3GPP}{3rd Generation Partnership Project}
\newacronym{5G}{5G}{fifth generation}
\newacronym{6G}{6G}{sixth generation}
\newacronym{bp}{BP}{belief propagation}
\newacronym{awgn}{AWGN}{additive white Gaussian noise}
\newacronym{ber}{BER}{bit error ratio}
\newacronym{bler}{BLER}{block error ratio}
\newacronym{cacfar}{CA-CFAR}{cell-averaging constant false alarm rate}
\newacronym{cn}{CN}{Core Network}
\newacronym{cqi}{CQI}{Channel Quality Indicator}
\newacronym{cfr}{CFR}{channel frequency response}
\newacronym{csi}{CSI}{channel state information}
\newacronym{dft}{DFT}{discrete Fourier transform}
\newacronym{dl}{DL}{downlink}
\newacronym{eq}{EQ}{equalization}
\newacronym{fr2}{FR2}{Frequency Range 2}
\newacronym{idft}{IDFT}{inverse Discrete Fourier transform}
\newacronym{isac}{ISAC}{integrated sensing and communication}
\newacronym{kpi}{KPI}{key performance indicator}
\newacronym{ldpc}{LDPC}{low-density parity-check}
\newacronym{llr}{LLR}{log-likelihood ratio}
\newacronym{los}{LoS}{line-of-sight}
\newacronym{mcs}{MCS}{modulation and coding scheme}
\newacronym{mo}{MO}{modulation order}
\newacronym{ofdm}{OFDM}{orthogonal frequency-division multiplexing}
\newacronym{prb}{PRB}{Physical Resource Block}
\newacronym{prs}{PRS}{Positioning Reference Signal}
\newacronym{qam}{QAM}{quadrature amplitude modulation}
\newacronym{qpsk}{QPSK}{quadrature phase-shift keying}
\newacronym{rcs}{RCS}{radar cross-section}
\newacronym{re}{RE}{Resource Element}
\newacronym{rf}{RF}{radio frequency}
\newacronym{rx}{RX}{receiver}
\newacronym{semf}{SeMF}{Sensing Management Function}
\newacronym{ser}{SER}{symbol error ratio}
\newacronym{snr}{SNR}{signal-to-noise ratio}
\newacronym{tx}{TX}{transmitter}
\newacronym{ue}{UE}{user equipment}
\newacronym{zf}{ZF}{zero-forcing}

\begin{document}

\title{Hybrid Resource Allocation Scheme for Bistatic ISAC with Data Channels} 

\author{
    \IEEEauthorblockN{
        Marcus Henninger,
        Lucas Giroto,
        Ahmed Elkelesh,
        Silvio Mandelli
        }

	\IEEEauthorblockA{Nokia Bell Labs Stuttgart, Germany
        \\
	E-mail:\{firstname.lastname\}@nokia-bell-labs.com}}

\maketitle

\begin{abstract}

Bistatic \gls{isac} enables efficient reuse of the existing cellular infrastructure and is likely to play an important role in future sensing networks. In this context,  \gls{isac} using the data channel is a promising approach to improve the bistatic sensing performance compared to relying solely on pilots. One of the challenges associated with this approach is resource allocation: the communication link aims to transmit higher \gls{mo} symbols to maximize the throughput, whereas a lower \gls{mo} is preferable for sensing to achieve a higher signal-to-noise ratio in the radar image. To address this conflict, this paper  introduces a hybrid resource allocation scheme: by placing lower \gls{mo} symbols as pseudo-pilots on a suitable sensing grid, we enhance the bistatic sensing performance while only slightly reducing the spectral efficiency of the communication link. Simulation results validate our approach against different baselines and provide practical insights into how decoding errors affect the sensing performance. 
\end{abstract}

\if\arxivdisclaimer1
\vspace{-0.5cm}
\blfootnote{\copyright~2026 IEEE. Personal use of this material is permitted. Permission from IEEE must be obtained for all other uses, in any current or future media, including reprinting/republishing this material for advertising or promotional purposes, creating new collective works, for resale or redistribution to servers or lists, or reuse of any copyrighted component of this work in other works.}
\else
\vspace{0.2cm}
\fi


\glsresetall

\section{Introduction}\label{sec:intro}

As one of the key new features of the upcoming \gls{6G} standard, \gls{isac} is set to equip cellular networks with radar-like functionalities \cite{gonzalez2024integrated}. While the most important use cases such as drone and intrusion detection have largely been agreed on \cite{ghosh2025unified, mandelli_isacsurvey} and research in both academia and industry has progressed rapidly in the last years, numerous challenges still remain to be addressed. Some of those are unique to bistatic \gls{isac} systems~\cite{giroto2026bistatic}, which offer great potential to reuse existing deployments and complement conventional monostatic setups.

In monostatic \gls{isac}, an estimate of the sensing \gls{csi} matrix can be readily obtained via \gls{zf} element-wise division of the received frame by the transmitted frame in the discrete frequency domain. 
In bistatic \gls{isac}, either only pilots or the full frame can be used to estimate the sensing \gls{csi} matrix~\cite{brunner2024bistatic}. In the former case, the typically sparse pilot allocation leads to a limited sensing performance due to the associated reduction of the available processing gain and maximum unambiguous range and Doppler shift capabilities. Increasing the allocation density to counteract this leads to an undesired reduction in spectral efficiency for the communication link. While using the full frame allows circumventing this, its implementation in practice is challenging since knowledge of the transmit symbols is typically not available at the \gls{rx}, requiring the data channel to be decoded.
However, also this results in a resource allocation conflict: sensing benefits from a low \gls{mcs} index (\iec low \gls{mo} and coding rate). This is due to the fact that lower \gls{mo} symbols are less prone to transmission errors than higher \gls{mo} symbols. As a consequence, they are still decodable at lower \gls{snr}, which improves the bistatic sensing performance. Additionally, constant amplitude constellation symbols have the further benefit that they do not lead to noise enhancement when computing the \gls{csi} matrix via \gls{zf}~\cite{braun2014ofdm} and exhibit lower sidelobes \cite{han2025constellation}. The communication link, however, aims at optimizing spectral efficiency and would thus prioritize selecting the highest possible \gls{mcs} to meet the target \gls{bler}.

The problem of bistatic sensing with the data channel has been investigated in several studies. In~\cite{gupta2025data}, refined estimates of the target parameters and data are obtained after a coarse estimation of the target parameters using a Bayesian learning approach. The authors of~\cite{keskin2025bridging} propose an iterative sensing and data demodulation technique, showing that the gap between pilot-only processing and a genie-aided scheme with perfect knowledge of the data symbols at the \gls{rx} can be narrowed. However, both works did not investigate the effects of channel coding present in practical systems. This aspect has been addressed in \cite{brunner2024bistatic}, where, among other aspects, the impact of decoding errors on the sensing performance has been studied. However, as in \cite{gupta2025data} and \cite{keskin2025bridging}, the previously introduced resource allocation problem between sensing and communication was only addressed to the extent that different pilot densities and \glspl{mo} were investigated. The use of superimposed pilots in \cite{bao2020superimposed} partially considers the aforementioned trade-off, but results in a power allocation problem involving a balance between pilots and information symbols. Furthermore, channel coding aspects were also not considered.

To close these gaps, we introduce a hybrid resource allocation scheme including channel coding for the data transmission, which addresses the previously introduced resource allocation conflict between communication and sensing. Specifically, we propose to allocate lower \gls{mo} pseudo-pilots on a suitable grid for sensing predefined by sensing \glspl{kpi}. Our main contributions are:
\begin{itemize}
    \item We propose a hybrid resource allocation scheme that facilitates bistatic sensing while keeping the impact on the communication link minimal.   
    \item Based on the proposed scheme, we detail the full processing pipeline to perform sensing at the bistatic \gls{rx}, including channel coding aspects.
    \item We show with simulations that our hybrid resource allocation scheme enhances the bistatic sensing performance. Moreover, our results reveal practical insights about the impact of decoding errors on the sensing performance.
\end{itemize}

\section{System Model}\label{sec:sys_model}

\begin{figure}[!t]
	\centering
	
	\psfrag{AAAA}[c][c]{\scriptsize $r^\text{TX---T}_p,\,f^\text{TX---T}_{\mathrm{D},p}$}
	\psfrag{BBBB}[c][c]{\scriptsize $r^\text{T---RX}_p,\,f^\text{T---RX}_{\mathrm{D},p}$}
	\psfrag{CCCC}[c][c]{\scriptsize $r_0,\,f_{\mathrm{D},0}$}
	\includegraphics[width=\columnwidth]{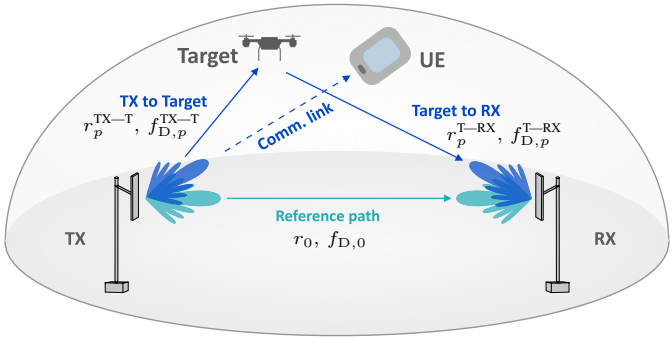}
	\captionsetup{justification=raggedright,singlelinecheck=false}
	\caption{Bistatic ISAC system model. In this example, a static reference path labeled as $p=0$ and a path $p=1$ associated with a radar target that is opportunistically illuminated by a transmission originally intended for a UE are shown. Here, \mbox{$r_p=r^\text{TX--T}_p+r^\text{T--RX}_p$} is the bistatic range resulting from the sum of the range $r^\text{TX--T}_p$ between TX and target and the range $r^\text{T--RX}_p$ between target and RX. In addition, a Doppler shift \mbox{$f_{\mathrm{D},p}=f^\text{TX---T}_{\mathrm{D},p}+f^\text{T---RX}_{\mathrm{D},p}$} is experienced.}
    \label{fig:sys_model}
	\vspace{-0.3cm}
\end{figure}

We consider the bistatic \gls{isac} system depicted in Fig.~\ref{fig:sys_model}. The \gls{tx} sends \gls{ofdm} symbols to a \gls{ue} in the \gls{dl} for communication purposes, while the \gls{rx}, \egc a receive-only node, processes reflected signals to gain knowledge about objects in the environment.%

The discrete-frequency domain \gls{tx} frame, which consists of~$M$ \gls{ofdm} symbols with~$N$ subcarriers that are spaced by $\Delta f$ and carry complex modulation symbols, is denoted by $\mathbf{X} \in \mathbb{C}^{N \times M}$ and transmitted at carrier frequency~$f_\text{c}$. The received frame $\mathbf{Y} \in \mathbb{C}^{N \times M}$ at the bistatic \gls{rx} is 
\begin{align}
\mathbf{Y} = \mathbf{X}\odot\mathbf{H} + \textbf{Z} \;,
\label{eq:Y}	
\end{align} 
where $\mathbf{Z} \in \mathbb{C}^{N \times M}$ is the random complex \gls{awgn} matrix, whose elements follow a Gaussian distribution with zero-mean and variance \mbox{$\sigma_{\text{n}}^2 = P_{\text{n}}/N$, with $P_{\text{n}}$} being the noise power over the whole bandwidth. We model the \gls{cfr} as a superposition of a strong path channel $\mathbf{H}_{p=0}$, \egc a \gls{los} path that can be used for synchronization and equalization, and a channel $\mathbf{H}_{\text{tar}}$ comprising weaker target paths indexed $p~\in~\mathcal{P}$ representing reflections due to objects in the surroundings. Thus, the \gls{cfr} writes as
\begin{align}
\mathbf{H} &= \mathbf{H}_{0} + \mathbf{H}_{\text{tar}} \nonumber \\ 
&= \alpha_{0} \cdot \mathbf{a}(r_0)\mathbf{b}(f_{\text{D},0})^\text{T} + \sum_{p \in \mathcal{P}} \alpha_p \cdot \mathbf{a} (r_p)\mathbf{b}(f_{\text{D},p})^\text{T} \;,
\label{eq:H}	
\end{align} 
where the channel contributions of a path with bistatic range~$r_p$ and bistatic Doppler shift $f_{\text{D},p}$ are described by the vectors
\begin{align}
\mathbf{a}(r_p) &= \begin{bmatrix}
       1, \ e^{-j2\pi \Delta f \cdot r_p/c_0}, \ \dots, \ e^{-j2\pi (N - 1) \Delta f \cdot r_p/c_0}
\end{bmatrix}^\text{T} \\
\mathbf{b}(f_{\text{D},p}) &= \begin{bmatrix}
       1, \ e^{j2\pi T_0 \cdot f_{\text{D},p}}, \ \dots, \ e^{j2\pi (M - 1) T_0  \cdot f_{\text{D},p}}
\end{bmatrix}^\text{T}  \; , 
\label{eq:channel_vectors}
\end{align}
where $T_0$ is the duration of an \gls{ofdm} symbol and $c_0$ denotes the speed of light. Note that in this work we assume perfect synchronization between \gls{tx} and \gls{rx} and negligible intercarrier interference (ICI) due to Doppler shifts. In practice, sufficiently accurate synchronization could be achieved, \egc by using a reference path as described in \cite{brunner2024bistatic}. Moreover, while we do not make assumptions about the angular domain in this work, angular processing (\egc beamforming) could be considered as described in \cite{felix2025}.
\section{Hybrid Resource Allocation Scheme}\label{sec:hybrid}

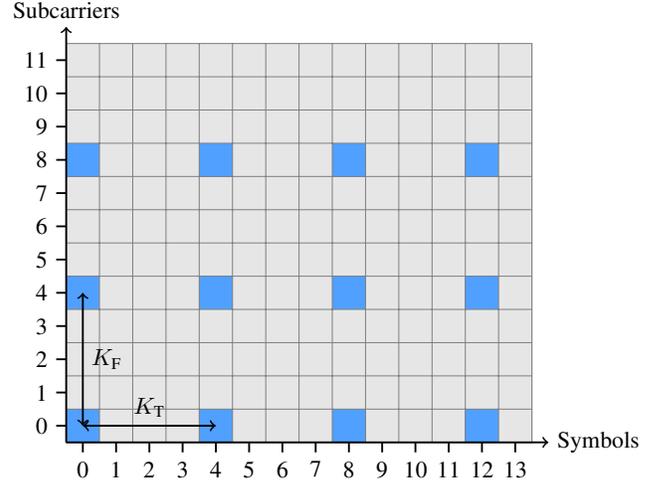
\begin{figure}
\centering
\resizebox{\linewidth}{!}{
\begin{tikzpicture}

\definecolor{grey}{RGB}{230, 230, 230}
\definecolor{blue}{RGB}{80, 160, 255}

\draw[black] (0,0) rectangle (7,6);
\fill[grey] (0,0) rectangle (7,6);

\foreach \x in {0, 2, ..., 6.5} {
    \foreach \y in {0, 2, ..., 5.5} {
        \fill[blue] (\x,\y) rectangle (\x+0.5,\y+0.5);
    }
}

\draw[step=0.5cm, gray, very thin] (0,0) grid (7,6);

\draw[->, thick] (0,0) -- (0,6.25) node[above] {Subcarriers};
\draw[->, thick] (0,0) -- (7.25,0) node[right] {Symbols};

\foreach \y in {0, 0.5, ..., 5.5} {
    \pgfmathsetmacro{\scaledy}{int(\y * 2)} 
    \draw[thick] (0,\y+0.25) -- (-0.15,\y+0.25) node[left] {\scaledy};
}
\foreach \x in {0, 0.5, ..., 6.5} {
    \pgfmathsetmacro{\scaledx}{int(\x * 2)} 
    \draw[thick] (\x+0.25,0) -- (\x+0.25,-0.15) node[below] {\scaledx};
}

\draw[<->, thick, black] (0.25,0.25) -- (2.25,0.25) node[midway, above, sloped] {$K_\text{T}$};

\draw[<->, thick, black] (0.25,0.25) -- (0.25,2.25) node[midway, right] {$K_\text{F}$};

\end{tikzpicture}
} 
\caption{Example of a hybrid resource allocation scheme, illustrated using one \acrshort{prb} spanning $N=12$ subcarriers and $M=14$ symbols. A lower \gls{mo} is used on every $K_\text{F} = K_\text{T} = 4$th \acrshort{re} (blue squares). The remaining grey \acrshortpl{re} use a higher \gls{mo}.}
  \label{fig:hybrid_scheme}
  \vspace{-0.3cm}
\end{figure}

\begin{figure*}[t]
\resizebox{\linewidth}{!}{
\begin{tikzpicture}[node distance=1cm]

\tikzset{every node/.style={font=\large}}%

\tikzstyle{emptyblock} = [rectangle, rounded corners, minimum width=1cm, minimum height=1.2cm, text centered, draw=white]
\tikzstyle{blocktx} = [rectangle, rounded corners, minimum width=3.3cm, minimum height=1.2cm, text centered, draw=black, fill=cyan!15]
\tikzstyle{blockrx} = [rectangle, rounded corners, minimum width=3.3cm, minimum height=1.2cm, text centered, draw=black, fill=orange!15]
\tikzstyle{process} = [rectangle, minimum width=2cm, minimum height=1.2cm, text centered, draw=black, fill=gray!20]
\tikzstyle{arrow} = [thick,->,>=stealth]
\tikzstyle{dashedboxcyan} = [draw=cyan, dashed, rounded corners, inner sep=1em]
\tikzstyle{dashedorangebox} = [draw=orange, dashed, rounded corners, inner sep=1em]

\node (encoding) [blocktx, xshift=-2cm] {Encoding};
\node (source) [emptyblock, left of=encoding, xshift=-2cm]{};
\node[align=center]  (modulationTX) [blocktx, right of=encoding, xshift=3cm] {Modulation and\\ RE Mapping};

\node (channel) [process, below of=modulationTX, xshift=3cm, yshift=-1cm] {Channel};

\node[align=center] (commsEQ) [blockrx, below of=channel, xshift =-3cm, yshift=-1cm] {Comm. EQ};
\node[align=center] (demodulation) [blockrx, left of=commsEQ, xshift=-3cm] {Demodulation and\\ RE Demapping};
\node (decoding) [blockrx, left of=demodulation, xshift=-3cm] {Decoding};
\node[align=center]  (modulationRX) [blockrx, left of=decoding, xshift=-3cm] {Modulation and\\ RE Mapping};
\node (sensingEQ) [blockrx, left of=modulationRX, xshift=-3cm] {Sensing EQ};
\node[align=center] (sensingProc) [blockrx, left of=sensingEQ, xshift=-3cm] {Sensing\\ Processing};
\node (sink) [emptyblock, left of=sensingProc, xshift=-3.8cm]{};

\draw[arrow] (source.east) -- node[anchor=south] {$\mathbf{b}$} (encoding.west);
\draw [arrow] (encoding) -- node[anchor=south] {$\mathbf{c}$} (modulationTX);
\draw [arrow] (modulationTX.east)  node[anchor=south, xshift=7.1mm] {$\mathbf{X}$} -- ++(1.35,0) --(channel.north);%

\draw [arrow] (channel.south) -- ++(0,-1.4) -- node[anchor=south] {$\mathbf{Y}$} (commsEQ);
\draw [arrow] (commsEQ) -- node[anchor=south] {$\mathbf{\hat{Y}}$} (demodulation);
\draw [arrow] (demodulation) -- node[anchor=south] {$\boldsymbol{\ell}$} (decoding);
\draw [arrow] (decoding) -- node[anchor=south] {$\mathbf{\hat{c}}$} (modulationRX);
\draw [arrow] (modulationRX) -- node[anchor=south] {$\mathbf{\hat{X}}$} (sensingEQ);
\draw [arrow] (sensingEQ) -- node[anchor=south] {$\mathbf{\hat{H}}$} (sensingProc);
\draw [arrow] (sensingProc.west) -- node[anchor=south, xshift = 0.2cm] {target list} (sink.east);

\draw [dashedboxcyan] ($(source.north west)+(0.7,0.35)$) rectangle ($(modulationTX.south east)+(0.35,-0.35)$);
\node[text=cyan] at ($(encoding.north west)+(-0.5,0.8)$) {TX};

\draw [dashedorangebox] ($(sensingProc.north west)+(-2.15,0.35)$) rectangle ($(commsEQ.south east)+(0.35,-0.35)$);
\node[text=orange] at ($(sensingProc.north)+(0,0.8)$) {Bistatic RX};

\end{tikzpicture}
} 
\caption{Processing pipeline for bistatic sensing.}
\label{fig:processing_pipeline}

\vspace{-0.3cm}

\end{figure*}
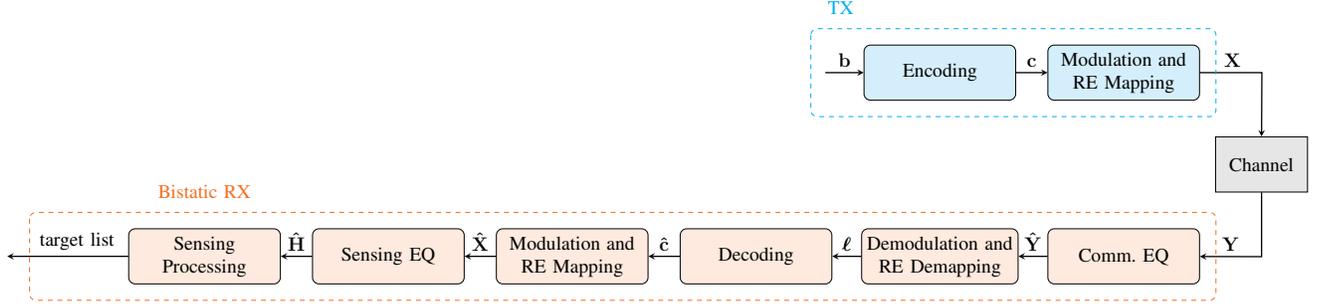

In the following, we detail our proposed hybrid resource allocation scheme, which aims to balance the trade-off between pilots and information symbols in an \gls{isac} context. The core idea is to not use a single, typically higher, \gls{mo}~$Q_\text{r}$ for the whole data channel, but to allocate symbols with a lower \gls{mo} $Q_\text{s}$ (\egc $Q_{\text{s}} = 2$ with \gls{qpsk}) as pseudo-pilots on a suitable grid for sensing. This enhances the sensing capability compared to using $Q_\text{r}$ for the entire frame, since lower \gls{mo} symbols become decodable ``earlier'', \iec at lower \gls{snr}. At the same time, the spectral efficiency reduction for the communication link is lower in comparison to using true dedicated pilots (\egc \gls{prs} symbols) that do not convey any information. An example of one \gls{prb} using the proposed scheme is shown in Fig.~\ref{fig:hybrid_scheme}, where the spacing between lower \gls{mo} symbols in frequency and time, denoted by $K_\text{F}$ and $K_\text{T}$, respectively, is~4.

The parameters of the hybrid allocation scheme can be determined based on both communication and sensing requirements. The former must be considered as in legacy resource allocation schemes to optimize communication performance by considering \glspl{kpi} like \gls{cqi} or rank indicator (RI) to choose the \gls{mo} $Q_\text{r}$ for the regular data symbols (grey \glspl{re} in Fig.~\ref{fig:hybrid_scheme}). The sensing requirements include a necessary sensing link budget and sensing \glspl{kpi}, \egc unambiguous range/Doppler shift or range/Doppler shift accuracy, and depend on the use case to be supported. In a \gls{3gpp}-compliant system, they will likely be communicated to the sensing system by the \gls{semf} in the \gls{cn}, which is foreseen to orchestrate the sensing operations. The parameters of the hybrid resource allocation scheme could be signalled to the \gls{ue} by extending existing \gls{mcs} tables, \egc based on \cite{3gpp_38214}.

A comprehensive analysis of how precisely the hybrid resource allocation scheme can be determined based on all the available inputs goes beyond the scope of this paper. Nonetheless, focusing only on the sensing requirements, we can formalize the following guidelines:
\begin{enumerate}
    \item \label{1} The required spacings $K_\text{F}$ and $K_\text{T}$ in frequency and time to support an unambiguous (bistatic) range $r_{\text{max}}'$ and Doppler shift~$f_{\text{D,max}}'$ can be obtained using the well-known formulae for the maximum achievable unambiguous range~$r_{\text{max}}$ and Doppler shift $f_{\text{D,max}}$ as follows~\cite{brunner2024bistatic}
    \begin{align}
        K_\text{F} &= \lfloor{\frac{r_{\text{max}}} {r_{\text{max}}'}\rfloor} = \lfloor{\frac{c_0}{\Delta f r_{\text{max}}'}\rfloor} \\   
        K_\text{T} &= \lfloor{\frac{f_{\text{D,max}}} {f_{\text{D,max}}'}\rfloor} = \lfloor{\frac{1}{2T_0 f_{\text{D,max}}'}\rfloor} \;,
    \end{align}
    where $\lfloor x \rfloor$ denotes the floor operator rounding down to the largest integer smaller than or equal to $x$.
    \item \label{2} Similarly, the required bandwidth $B$ and duration $T_\text{B}$ of the sensing burst to enable a range resolution $\Delta r'$ and Doppler shift resolution $\Delta f_{\text{D}}'$ can be computed based on the formulae for the  range resolution $\Delta r$ and Doppler shift resolution $\Delta f_{\text{D}}$ of the system~\cite{brunner2024bistatic}
    \begin{align}
    B &= 
    \frac{c_0} {\Delta r'} \\  
    T_\text{B} &= 
    \frac{1} {\Delta f_{\text{D}}'} \; .
    \end{align}
    \item Finally, link budget requirements should be taken into account. If the sensing burst allocation determined based on \ref{1}) and \ref{2}) is not sufficient to support the maximum sensing range required by the use case, lower \gls{mo} symbols should be allocated to additional resources to increase the processing gain.
\end{enumerate}
\begin{remark}
    The previous guidelines are formulated based on the simplifying assumption that all lower \gls{mo} pseudo-pilots will be correctly decoded at the bistatic \gls{rx}, while the regular data symbols are not correctly decoded. Even though this clearly does not hold in practice and is also not the case in our simulations later, it can serve as an initial assumption that can later be refined, \egc by observing the resulting sensing \glspl{kpi} based on the used sensing burst allocation.
\end{remark}
The resulting spectral efficiency per symbol with our hybrid resource allocation scheme as defined in \cite{3gpp_38214} is given as 
\begin{align}
\eta = \frac{R \left(N_{\text{s}}^{\text{RE}} Q_{\text{s}}+ \left(NM-N_{\text{s}}^{\text{RE}}\right)Q_{\text{r}} \right)}{NM} \;, \label{eq:spec_eff}
\end{align}
where $N_{\text{s}}^{\text{RE}} =\lceil{N/K_{\text{F}}}\rceil\cdot \lceil{M/K_{\text{T}}}\rceil$ is the number of resource elements on the sensing grid, with $\lceil x \rceil$ denoting the ceil operator rounding up to the smallest integer greater than or equal to $x$. Moreover, $R$ is the code rate, which for simplicity is assumed to be the same for the sensing grid and regular communication symbols.

\section{Bistatic Sensing Processing}\label{sec:bistatic_pipeline}

Next, we describe the full processing pipeline shown in 
Fig.~\ref{fig:processing_pipeline}. Performing these operations allows us to extract sensing information at the bistatic sensing \gls{rx} using our proposed hybrid resource allocation scheme. The pipeline comprises the following steps:
\\
\fakepar{Encoding} The information bitstream vector $\mathbf{b}$ to be transmitted is encoded into the codeword vector $\mathbf{c}$.
\\ \fakepar{Modulation and RE Mapping} The codeword vector is modulated to constellation symbols using the modulation alphabets according to $Q_{\text{r}}$ and $Q_{\text{s}}$. The resulting symbols are then mapped to the \glspl{re} on the grid to obtain the \gls{tx} frame~$\mathbf{X}$.%
\\
\fakepar{Channel} The \gls{tx} frame $\mathbf{X}$ is transmitted over the channel, which we model as described in Sec.~\ref{sec:sys_model}.
\\
\fakepar{Comm. EQ} Before decoding the communication signals, the \gls{rx} frame $\mathbf{Y}$ must be equalized. As \gls{eq} is not the focus of this work, we did not implement any channel estimation techniques based on dedicated pilot transmissions. 

Instead, we assume the presence of an \gls{los} path. Based on~\cite{brunner2024bistatic}, this enables accurate synchronization and therefore compensation of carrier frequency offset and the experienced Doppler shift in this path. Consequently, intercarrier interference is avoided and estimation of the channel of the dominant path $\mathbf{H}_{0}$ at the \gls{rx} is facilitated. Perfect knowledge of $\mathbf{H}_{0}$ and $\mathbf{H}_0 \approx \mathbf{H}$ are henceforth considered for communication, which is reasonable as a dominant \gls{los} path is assumed. The \gls{rx} frame after \gls{eq} with matrix $\mathbf{G}$ then writes as
\begin{align}
\hat{\mathbf{Y}} = \mathbf{Y}\odot\mathbf{G} \;,
\label{eq:comms_eq}
\end{align}
where \gls{zf} with the dominant path is applied. Thus, 
\begin{align}
\left[\mathbf{G}\right]_{n, m} = \frac{1}{\left[\mathbf{H}_{0}\right]_{n, m}} \;,
\label{eq:comms_eq_zf}
\end{align}
with $n$ and $m$ denoting subcarrier and \gls{ofdm} symbol index, respectively. 
\\
\fakepar{Demodulation and RE Demapping} The \gls{rx} frame after \gls{eq} is first serialized by demapping it from the \gls{re} grid. After that, the resulting symbols are demodulated by computing the \gls{llr} values $\boldsymbol{\ell}$. 
\\
\fakepar{Decoding} Hard estimates of the codeword bits $\mathbf{\hat{c}}$ are obtained by decoding the \gls{llr} values $\boldsymbol{\ell}$.
\\
\fakepar{Modulation and RE Mapping} Before performing further sensing processing, the estimated codeword vector $\mathbf{\hat{c}}$ is modulated to constellation symbols and then again mapped to the \gls{re} grid to obtain the matrix containing the estimated symbols~$\mathbf{\hat{X}}$. Note that this step corresponds to the step ``Modulation and RE Mapping'' performed at the \gls{tx}.
\\
\fakepar{Sensing EQ} 
To obtain an estimate of the full channel for sensing, the transmitted symbols are equalized via \gls{zf} using the estimated symbols as
\begin{align}
\left[\hat{\mathbf{H}}\right]_{n, m} = \frac{\left[\hat{\mathbf{Y}}\right]_{n, m}}{\left[\hat{\mathbf{X}}\right]_{n, m}} \;.
\label{eq:sensing_eq}
\end{align}
\\
\fakepar{Sensing Processing} 
Finally, the \gls{csi} matrix $\hat{\mathbf{H}}$ is processed to extract sensing information. We employ conventional 2D \gls{dft} processing to obtain the periodogram (range-Doppler radar image) $\mathbf{P}$ as~\cite{braun2010maximum} 
\begin{align}
\left[\mathbf{P}\right]_{n, m}= \frac{1}{N'M'}\bigg|\sum_{k=0}^{N'-1} \Biggl(\sum_{l=0}^{M'-1} \hat{\mathbf{H}}(k, l)e^{-j2\pi\frac{lm}{M'}}\Biggr)e^{j2\pi\frac{kn}{N'}} \bigg|^2
\label{eq:per} \; ,
\end{align}
where $N'$ and $M'$ denote the number of rows and columns of~$\hat{\mathbf{H}}$ after zero padding. Note that despite the approximation $\mathbf{H}_0 \approx \mathbf{H}$, the processing gain enhances the \gls{snr} of the paths due to the focusing of the \gls{dft} operations in \eqref{eq:per}, thus also facilitating the detection of weaker target contributions in $\mathbf{H}_{\text{tar}}$ in the periodogram. Subsequently, a \gls{cacfar} detector~\cite{richards2010principles} is applied to obtain a list of target peaks, each characterized by a range and Doppler shift estimate.
\section{Results and Discussion}\label{sec:results}

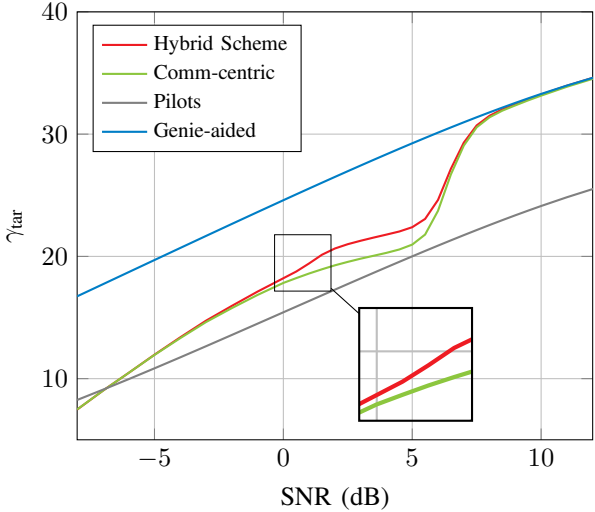
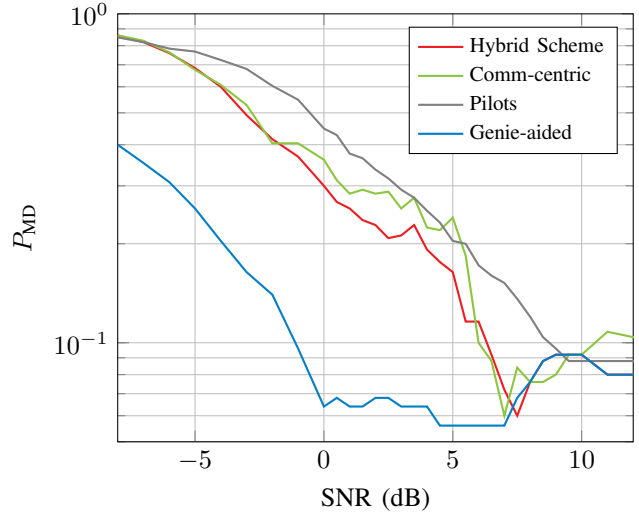
\begin{figure*}[t!]
\centering
\captionsetup[subfigure]{labelsep=space}
\begin{subfigure}{0.49\textwidth}
    \def\scale{1}

\begin{tikzpicture}
        [spy using outlines={rectangle, magnification=2, size=1cm, connect spies}] 
        
		\begin{axis}[
			height = 7cm,
    		xlabel={SNR (dB)},
    		ylabel={$\gamma_{\text{tar}}$ (dB)},
    		xmin = -8,
            xmax = 12,
    		ymin=5,
		    ymax = 40,
            ytick={10, 20, 30, 40, 50},
		    enlargelimits = false,
    		xmajorgrids=true,
    		ymajorgrids=true,
    		grid style=solid,
    		legend pos = north west,
		    legend style={font=\footnotesize},
            legend columns = 1,
            transpose legend,
            legend cell align={left},
			every axis plot/.append style={thick},
        	scale = \scale,
            scaled ticks=false, 
            mark repeat={2}
		]

        \addplot[color=rot, thick]
    	plot table[x expr=\thisrowno{0}, y index=2] {data/Image_SNR_Inf_Vec_Len512_CW_len_1024_N_weak_5_Spacing_freq_4_Spacing_time_4.txt};

        \addlegendentry{Hybrid Scheme}

        \addplot[color=apfelgruen,thick]
    	plot table[x expr=\thisrowno{0}, y index=4] {data/Image_SNR_Inf_Vec_Len512_CW_len_1024_N_weak_5_Spacing_freq_4_Spacing_time_4.txt};

        \addlegendentry{Comm-centric}

        \addplot[color=mittelgrau,thick]
    	plot table[x expr=\thisrowno{0}, y index=3] {data/Image_SNR_Inf_Vec_Len512_CW_len_1024_N_weak_5_Spacing_freq_4_Spacing_time_4.txt};

        \addlegendentry{Pilots}

        \addplot[color=mittelblau, thick]
    	plot table[x expr=\thisrowno{0}, y index=1] {data/Image_SNR_Inf_Vec_Len512_CW_len_1024_N_weak_5_Spacing_freq_4_Spacing_time_4.txt};

        \addlegendentry{Genie-aided}

	\end{axis}

\spy[black,size=1.25cm] on (3,2.25) in node [fill=none] at (4.5, 1);

\end{tikzpicture}
    \caption{Target SNR}
   \label{fig:target_snr}
\end{subfigure}
\begin{subfigure}{0.49\textwidth}
\def\scale{1}

\begin{tikzpicture}
        [spy using outlines={rectangle, magnification=2.5, size=1cm, connect spies}] 
        
		\begin{semilogyaxis}[
			height = 7cm,
    		xlabel={SNR (dB)},
    		ylabel={$P_{\text{MD}}$},
    		xmin = -8,
            xmax = 12,
    		ymin=0.05,
		    ymax = 1,
		    enlargelimits = false,
    		xmajorgrids=true,
    		yminorgrids=true,
    		grid style=solid,
    		legend pos = north east,
		    legend style={font=\footnotesize},
            legend columns = 1,
            transpose legend,
            legend cell align={left},
			every axis plot/.append style={thick},
        	scale = \scale,
            mark repeat={2}
		]

        \addplot[color=rot, thick]
    	plot table[x expr=\thisrowno{0}, y index=2] {data/Missed_Det_Inf_Vec_Len512_CW_len_1024_N_weak_5_Spacing_freq_4_Spacing_time_4.txt};

        \addlegendentry{Hybrid Scheme}

        \addplot[color=apfelgruen,thick]
    	plot table[x expr=\thisrowno{0}, y index=4] {data/Missed_Det_Inf_Vec_Len512_CW_len_1024_N_weak_5_Spacing_freq_4_Spacing_time_4.txt};

        \addlegendentry{Comm-centric}

        \addplot[color=mittelgrau,thick]
    	plot table[x expr=\thisrowno{0}, y index=3] {data/Missed_Det_Inf_Vec_Len512_CW_len_1024_N_weak_5_Spacing_freq_4_Spacing_time_4.txt};

        \addlegendentry{Pilots}

        \addplot[color=mittelblau, thick]
    	plot table[x expr=\thisrowno{0}, y index=1] {data/Missed_Det_Inf_Vec_Len512_CW_len_1024_N_weak_5_Spacing_freq_4_Spacing_time_4.txt};

        \addlegendentry{Genie-aided}
                               
	\end{semilogyaxis}

\end{tikzpicture} 
    \caption{Probability of missed detection}
   \label{fig:pmd}
\end{subfigure}
\caption{Target SNR in the periodogram and probability of missed detection for our proposal ``Hybrid Scheme'' (red), ``Comm-centric'' (green), ``Pilots'' (grey), and ``Genie-aided'' (blue), both as functions of the receive SNR before radar processing.}
\label{fig:sensing_stats}

\vspace{-0.3cm}

\end{figure*}

\subsection{Bistatic Sensing Scenario}

To assess the capability of our proposed hybrid resource allocation scheme in conjunction with the previously described bistatic sensing processing steps, we run Monte Carlo simulations based on the system model defined in Sec. \ref{sec:sys_model}. The simulation parameters are summarized in Table~\ref{tab:params}.

\begin{table}[t]
    \vspace{0.2cm}
    \caption{Simulation parameters. \label{tab:params}}
	\centering
     \begin{tabu}{|l|r|}
            \hline
            \textbf{Parameter} & \textbf{Value} \\
			\Xhline{3\arrayrulewidth}
			Carrier frequency $f_\text{c}$ & 27.4 GHz \\
  			\hline
  			Number of subcarriers $N$ & 792 \\
  			\hline
  			Subcarrier spacing $\Delta f$ & 120 kHz \\
  			\hline
            Total bandwidth $B$ & 95 MHz \\
            \hline
            Number of symbols $M$ per frame & 560 \\
  			\hline
            Spacing in frequency $K_\text{F}$ & 4 \\
  			\hline
            Spacing in time $K_\text{T}$ & 4 \\
  			\hline
            \gls{mo} for sensing grid symbols $Q_{\text{s}}$ & 2 (QPSK) \\
            \hline
            \gls{mo} for regular data symbols $Q_{\text{r}}$ & 4 (16-QAM) \\
            \hline
            Code rate $R$ & 0.5 \\
            \hline
            Max. iterations for BP decoding & 20 \\
            \hline
            \acrshort{cacfar} prob. of false alarm & $10^{-5}$ \\
            \hline
    \end{tabu}
\end{table}

In each realization, we place the dominant path at a random bistatic range $r_0$ between 200 and 300~m, which are typical distances in practical deployments \cite{giroto2026bistatic}. For the $\left| \mathcal{P}\right| = 5$ point targets to be detected at the bistatic \gls{rx}, excess ranges between 30 and 90~m w.r.t. the reference path are considered. This results in ranges $r_p$ between 230 and 390~m and, disregarding \gls{rcs}, amounts to path losses for radar targets between approximately 47 and 53 dB higher than that of the reference path. All paths are associated with a random Doppler shift $f_{\text{D}}$, with the maximum possible Doppler shift being $1.8$~kHz. For comparison, this is the Doppler shift that would result from a radial velocity of 10 m/s if monostatic sensing were performed. The dominant reference path is excluded from the evaluation of the sensing performance, as we assume it to be perfectly known for \gls{eq}. Moreover, all random complex coefficients $\alpha$ include attenuation due to free space path loss.

Assuming a \gls{fr2} setup with carrier frequency $f_{\text{c}}$ and subcarrier spacing $\Delta f=120$~kHz, each sensing frame comprises $N=792$ subcarriers and $M=560$ symbols. For our proposed hybrid resource allocation scheme, we use the configuration depicted in Fig.~\ref{fig:hybrid_scheme}, \iec with spacings $K_{\text{F}} = K_{\text{T}} = 4$ and \gls{qpsk} modulation ($Q_{\text{s}}=2$) on the sensing grid. The remaining symbols employ 16-\gls{qam} modulation ($Q_{\text{r}}=4$). Furthermore, we use the Sionna framework~\cite{hoydis2022sionna} for channel coding. The \gls{5G}-compliant \gls{ldpc} code has a code rate $R=0.5$ and is decoded with a maximum of 20 \gls{bp} iterations with an early stopping criterion at the \gls{rx}. According to \eqref{eq:spec_eff}, the hybrid allocation scheme parametrized in this way achieves a spectral efficiency of $\eta_{\text{hybrid}} = 1.9375$ bits per symbol.

We evaluate the sensing performance by sweeping the \gls{snr} at the bistatic \gls{rx} and examining the probability of missed detection $P_{\text{MD}}$ and target \gls{snr}~$\gamma_ {\text{tar}}$. The former is the percentage of undetected targets, while we define the latter as  
\begin{align}
\gamma_ {\text{tar}} = \frac{\sum_{p \in \mathcal{P}} \left[\mathbf{P}\right]_{k_p, l_p}}{P_\text{res}} \; ,
\label{eq:target_snr}
\end{align}
\iec as the ratio between the mean power of the bins with the maximum target peak contribution (\iec one bin for each target peak) and the residual noise power $P_\text{res}$. In \eqref{eq:target_snr}, $k_p$ and $l_p$ are the range and Doppler indices of the $p$-th target peak in the periodogram, obtained using ground truth information. Moreover, $P_\text{res}$ is computed by averaging the periodogram power excluding bins associated with the main lobes of the paths. Overall, we simulate 50 realizations per \gls{snr} point. While this may seem like a relatively low number, it implies decoding ca. 85000 codewords of length~1024, which is a large enough number to gain meaningful insights into the influence of decoding errors on the sensing performance (down to a communication \gls{bler} of~$10^{-3}$).

We compare our proposal to the following baselines: \\
\fakepar{Comm-centric} The communication-centric baseline uses 16-\gls{qam} with $Q_\text{r} = 4$ for the whole data channel, \iec without interleaving lower \gls{mo} symbols on the sensing grid. Thus, it aims at achieving a high spectral efficiency for the communication link without considering sensing requirements. \\
\fakepar{Pilots} Sensing is performed only based on true \gls{qpsk} pilots that are allocated on the  same grid, \iec with $K_{\text{F}} = K_{\text{T}} = 4$. \\
\fakepar{Genie-aided} The third baseline uses the same hybrid allocation scheme as our proposal, but assumes knowledge of the \gls{tx} data at the bistatic \gls{rx}, \iec no decoding is necessary.

\subsection{Bistatic Sensing Performance}

In the following, we restrict our evaluation to the sensing performance at the bistatic \gls{rx}, \iec we do not model and evaluate the throughput of the communication link. 

Fig.~\ref{fig:target_snr} displays the target \gls{snr} in the periodogram $\gamma_{\text{tar}}$ for our proposed hybrid scheme and the three baselines. First, one can see that ``Pilots'' exhibits the worst performance over almost the whole \gls{snr} range before radar processing, which further underscores the need for sensing with the data channel to enhance the processing gain. The baseline ``Genie-aided'' with knowledge of the \gls{tx} symbols for sensing \gls{eq} expectedly achieves the best performance. A close comparison of ``Hybrid Scheme'' and ``Comm-centric'' reveals the advantages of our proposal: around an \gls{snr} of $0$~dB, a target \gls{snr} gap of up to $1.5$~dB emerges between ``Hybrid Scheme'' and ``Comm-centric''. This behavior can be attributed to the \gls{qpsk} symbols we allocate on the sensing grid, which become decodable at a lower \gls{snr} compared to the 16-\gls{qam} symbols that ``Comm-centric'' is using for the entire data channel. The pronounced gap extends over an \gls{snr} range of ca. $6$~dB, which corresponds to the expected gap between the waterfall regions for decoding \gls{qpsk} and 16-\gls{qam}. For higher \gls{snr}, both ``Hybrid Scheme'' and ``Comm-centric'' converge to ``Genie-aided'', as all symbols can be decoded virtually without errors.

While the probability of missed detection curves in Fig.~\ref{fig:pmd} are less smooth compared to Fig.~\ref{fig:target_snr} due to the limited number of detections per \gls{snr} point, they show that the improved~$\gamma_{\text{tar}}$ translates to an enhanced detection capability and confirm the previously observed trends. An improved target detection capability of ``Hybrid Scheme'' compared to ``Comm-centric'' is observed in particular in the higher \gls{snr} region.

Nonetheless, it should be noted that the improvements come at the cost of a spectral efficiency reduction of ca. $3$\% (with $\eta_{\text{hybrid}} = 1.9375$ and $\eta_{\text{comm}} = 2$ bits per symbol) for the communication link. Higher gains could be achieved by using $Q_{\text{r}} > 4$ or if the lower \gls{mo} symbols would be allocated more densely on the sensing grid by reducing $K_{\text{F}}$ and $K_{\text{T}}$. Also reducing the code rate for the lower \gls{mo} symbols would protect them more strongly and thus benefit sensing. However, exploring these degrees of freedom to aid symbol detection at the bistatic \gls{rx} would result in a further spectral efficiency reduction for the communication link. The parameter choice in practice thus requires careful consideration of the aspects described in Sec.~\ref{sec:hybrid} to trade off communication performance vs. sensing performance.

\subsection{Impact of decoding errors on sensing performance}

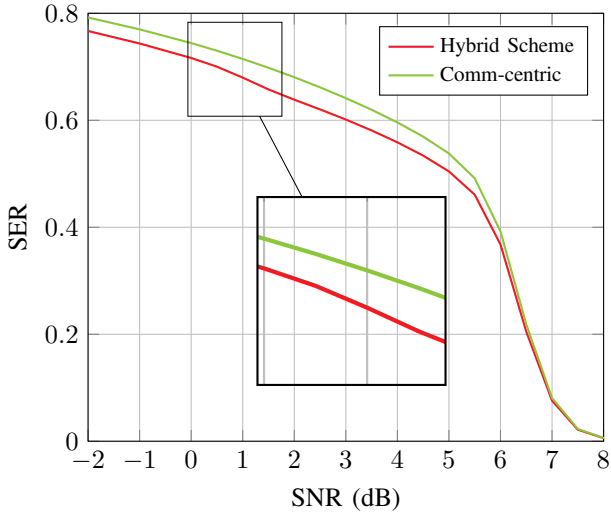
\begin{figure}[t!]
\centering
\def\scale{1}

\begin{tikzpicture}
        [spy using outlines={rectangle, magnification=2, size=1cm, connect spies}] 
        
		\begin{semilogyaxis}[
			height = 7cm,
    		xlabel={SNR (dB)},
    		ylabel={SER},
    		xmin = -2,
            xtick ={-2, -1, ..., 10},
            ytick ={1, 0.1, 0.01, 0.001},
            xmax = 10,
    		ymin= 0.001,
		    ymax = 1,
		    enlargelimits = false,
    		xmajorgrids=true,
    		ymajorgrids=true,
    		grid style=solid,
    		legend pos = south west,
		    legend style={font=\footnotesize},
            legend columns = 1,
            transpose legend,
            legend cell align={left},
			every axis plot/.append style={thick},
        	scale = \scale,
            mark repeat={2}
		]

        \addplot[color=rot, thick]
    	plot table[x expr=\thisrowno{0}, y index=1] {data/SER_Inf_Vec_Len512_CW_len_1024_N_weak_5_Spacing_freq_4_Spacing_time_4.txt};

        \addlegendentry{Hybrid Scheme}

        \addplot[color=apfelgruen,thick]
    	plot table[x expr=\thisrowno{0}, y index=2] {data/SER_Inf_Vec_Len512_CW_len_1024_N_weak_5_Spacing_freq_4_Spacing_time_4.txt};

        \addlegendentry{Comm-centric}
                               
	\end{semilogyaxis}

\spy[black,width=2.5cm, height=1cm] on (1.72, 5.1) in node [fill=none] at (1.72, 3.5);

\spy[black,width=1.25cm, height=1.5cm] on (4.33, 4.6) in node [fill=none] at (4.33, 1.8);

\end{tikzpicture}
\caption{\Acrfull{ser} at the bistatic \gls{rx} for our proposal ``Hybrid Scheme'' (red) and ``Comm-centric'' (green) as function of the receive SNR before radar processing.}
\label{fig:ser}

\vspace{-0.3cm}

\end{figure}

To get further insights into the impact of data decodability for sensing, we plot the \gls{ser} for the approaches that require decoding at the bistatic \gls{rx}, \iec our proposed hybrid scheme and the communication-centric baseline, in Fig.~\ref{fig:ser}. We choose the \gls{ser} as a metric over the \gls{ber}/\gls{bler} metrics typically used in communication, as the \gls{ser} at the bistatic \gls{rx} is directly coupled to the quality of the sensing \gls{eq} and thus the quality of the \gls{csi} estimation for sensing.

One can observe that ``Hybrid Scheme'' achieves a lower \gls{ser} than ``Comm-centric'' due to the \gls{qpsk} symbols on the sensing grid. In line with Fig.~\ref{fig:sensing_stats}, the gap widens between ca. $0$ and $1.5$~dB~\gls{snr}, which corresponds to the waterfall region where the \gls{qpsk} symbols become decodable. The seemingly minor impact on the \gls{ser} curves in Fig.~\ref{fig:ser} is explained by the relatively low share of \gls{qpsk} symbols, amounting to only $6.25$\% of all symbols. The gap closes at around $6$~dB when also the 16-\gls{qam} symbols become decodable.

Together with the results from Fig.~\ref{fig:sensing_stats}, the \gls{ser} curves allow us to draw the following conclusions on the impact of decoding errors on the sensing performance: sensing using the data channel does not necessarily require correct decoding of the entire frame. Rather, each correctly decoded symbol benefits the \gls{csi} estimation, comparable to a processing gain due to more available \glspl{re}. Decoding errors therefore appear to have no direct detrimental effect on sensing apart from a reduced processing gain, as long as they are random and do not exhibit specific patterns as discussed in~\cite{brunner2024bistatic}.
\section{Conclusion}\label{sec:conclusion}

In this paper, we proposed a hybrid resource allocation scheme to facilitate bistatic sensing operations using data channels. By allocating lower \gls{mo} symbols as pseudo-pilots on a suitable grid for sensing, we facilitate the correct decoding of the data channel symbols at the bistatic \gls{rx}. Our simulation results show that using the resulting improved \gls{csi} matrix estimation for sensing leads to a higher target \gls{snr} in the radar image accompanied by an improved detection performance compared to the communication-centric baseline. Moreover, we gained insights about the impact of the decoding errors on the sensing performance.
Future work should include a more extensive exploration of the various degrees of freedom for the hybrid resource allocation scheme and a holistic investigation that also considers the performance of the communication link.

\section*{Acknowledgments}
The authors acknowledge the financial support by the Federal Ministry of Research, Technology and Space of Germany in the project “SENSATION” under grant number 16KIS2523K.

\bibliographystyle{IEEEtran}
\bibliography{hybrid}

\end{document}